\documentclass[12pt]{iopart}

\usepackage{graphicx}
\usepackage{subfig}
\usepackage{amssymb}
\usepackage{amsthm} 
\usepackage{mathptmx}
\usepackage{nicefrac}
\usepackage{xcolor}
\begin{document}

\title[]{The SIDDHARTA-2 calibration method for high precision kaonic atoms X-ray spectroscopy measurements}

\author{F Sgaramella$^{1*}$, M Miliucci$^{1**}$, M Bazzi$^1$, D Bosnar$^2$, M Bragadireanu$^3$, M Carminati$^4$, M Cargnelli$^5$, A Clozza$^1$, G Deda$^4$, L De Paolis$^1$, R Del Grande$^{1,6}$, C Fiorini$^4$, C Guaraldo$^1$, M Iliescu$^1$, M Iwasaki$^7$, P King$^4$, P Levi Sandri$^1$, J Marton$^5$, P Moskal$^8$, F Napolitano$^1$, S Nied\'{z}wiecki$^8$, K Piscicchia$^{9,1}$, A Scordo$^1$, H Shi$^5$, M Silarski$^8$, D Sirghi$^1$, F Sirghi$^1$, M Skurzok$^8$, A Spallone$^1$, M T\"uchler$^5$, J Zmeskal$^5$ and C Curceanu$^1$ }

\address{$^1$ INFN-LNF, Istituto Nazionale di Fisica Nucleare-Laboratori Nazionali di Frascati, Frascati, 00044 Roma, Italy}
\address{$^2$ Department of Physics, Faculty of Science, University of Zagreb, 10000 Zagreb, Croatia}
\address{$^3$ Horia Hulubei National Institute of Physics and Nuclear Engineering IFIN-HH Măgurele, Romania}
\address{$^4$ Politecnico di Milano, Dipartimento di Elettronica, Informazione e Bioingegneria and INFN Sezione di Milano, 20133 Milano, Italy}
\address{$^5$ Stefan-Meyer-Institut f\"ur Subatomare Physik, 1090 Vienna, Austria}
\address{$^6$ Excellence Cluster Universe, Technische Universit\"{a}t M\"unchen, 85748 Garching, Germany}
\address{$^7$ RIKEN, Tokyo 351-0198, Japan}
\address{$^8$ The M. Smoluchowski Institute of Physics, Jagiellonian University, 30-348 Krak\'ow, Poland}
\address{$^9$ Centro Ricerche Enrico Fermi—Museo Storico della Fisica e Centro Studi e Ricerche “Enrico Fermi”, 00184 Roma, Italy}

\ead{$^*$ francesco.sgaramella@lnf.infn.it (Corresponding Author)}
\ead{$^{**}$ marco.miliucci@lnf.infn.it (Corresponding Author)}

\vspace{10pt}
\begin{indented}
\item[] January 2022
\end{indented}

\begin{abstract}
The SIDDHARTA-2 experiment at the DA$\Phi$NE collider aims to perform the first kaonic deuterium X-ray transitions to the fundamental level measurement, with a systematic error at the level of a few eV. To achieve this challenging goal the experimental apparatus is equipped with 384 Silicon Drift Detectors (SDDs) distributed around its cryogenic gaseous target. The SDDs developed by the SIDDHARTA-2 collaboration are suitable for high precision kaonic atoms spectroscopy, thanks to their high energy and time resolutions combined with their radiation hardness. The energy response of each detector must be calibrated and monitored to keep the systematic error, due to processes such as gain fluctuations, at the level of 2-3 eV. This paper presents the SIDDHARTA-2 calibration method which was optimized during the preliminary phase of the experiment in the real background conditions of the DA$\Phi$NE collider, which is a fundamental tool to guarantee the high precision spectroscopic performances of the system over long periods of data taking, as that required for the kaonic deuterium measurement.
\end{abstract}

%
\vspace{2pc}
%
\submitto{\PS}
%
%
%

\section{Introduction}
The study of the strong interaction is the main motivation for the exotic hadronic atoms experiments. Among the exotic atoms experiments, the kaonic atoms spectroscopy plays a special role, since it allows to obtain direct information about the strong interaction in the strangeness sector in the non-perturbative regime. Within this framework, the progress in the X-ray detection technology has been fundamental, leading to the realization of high spectroscopic performance devices, able to operate in the high background present in particle colliders and accelerators. Over the years, the detector technology improvements allowed to obtain an increase of the signal-to-background ratio and, consequently, to perform even more precise measurements. The main advantage in the use of semiconductor X-ray detectors lies in the fact that the energy required to create electron-hole pairs is very low compared, for example, to a gas detector, leading to a higher energy resolution. The breakthrough in the development of silicon detectors for X-ray spectroscopy was the introduction of the Silicon Drift Detectors (SDDs). SDDs, developed from the Silicon Drift Chamber technology introduced by Gatti and Rehak \cite{Gatti:1983wdm,Gatti:1984zk}, are now used for high precision X-ray spectroscopy thanks to their high energy resolution, rate capability and timing information.\\
Table \ref{tab1} shows a comparison of the main characteristics of the silicon detectors used for X-ray spectroscopy. The Si(Li) detector \cite{Iwasaki:1997wf} and the Charge-Coupled Device (CCD) \cite{Ishiwatari:2006yb} were typically used for kaonic atom experiments, but then the application of SDDs for X-ray spectroscopy allowed to perform experiments much more accurately with respect to the past, thanks to their high energy and time resolutions. SDDs were firstly used for X-ray spectroscopy in the E570 experiment at KEK 12 GeV-PS to solve the ``kaonic helium-4 puzzle" \cite{Okada:2007ky}. Subsequently, the SIDDHARTA experiment at INFN-LNF employed these detectors to perform the most precise measurement of the kaonic hydrogen \cite{SIDDHARTA:2011dsy} and also the first measurement of kaonic helium-3 \cite{SIDDHARTA:2010uae}. New Silicon Drift Detectors (SDDs-CUBE) have been developed specifically for the SIDDHARTA-2 experiment \cite{Curceanu:2019uph}, with improved performances with respect to the SDD-JFET (see Section \ref{sec_sdd}) used for the SIDDHARTA experiment, with the aim to perform the challenging and unprecedented kaonic deuterium measurement. In this paper we report the spectroscopic response of the SIDDHARTA-2 SDDs, focusing on the optimization of the calibration method of the device, as a fundamental tool for the kaonic deuterium measurement. 

\begin{table}[ht]	
\begin{center}
	\begin{tabular}{ccccc} 
		\hline\hline
		 \textbf{Detector} & \textbf{Si(Li)}\cite{Iwasaki:1997wf} & \textbf{CCD}\cite{Ishiwatari:2006yb} & \textbf{SDD-JFET}\cite{LECHNER1996346} & \textbf{SDD-CUBE}\cite{Miliucci:2019mdpi}\\ 
		 \hline 
		 Effective area (mm$^2$)                &200&724&3 x 100&8 x 64\\
		 Thickness (mm)                         &5   &0.03&0.45     &0.45\\
		 Energy resolution (eV) at 6 keV    &410&150&160       &140\\
		 Drift time (ns)                                 &290&---  &800       &400\\
		  Experiment                                    &KpX&DEAR&SIDDHARTA&SIDDHARTA-2, E57\\

		 \hline\hline
	\end{tabular}
\caption{Comparison between silicon detectors for X-ray spectroscopy.}
	\label{tab1}
	\end{center}
\end{table}

\section{The Silicon Drift Detectors}\label{sec_sdd}

\subsection{Working principle}
\begin{figure}[htbp]
\centering
\mbox{\includegraphics[width=11 cm]{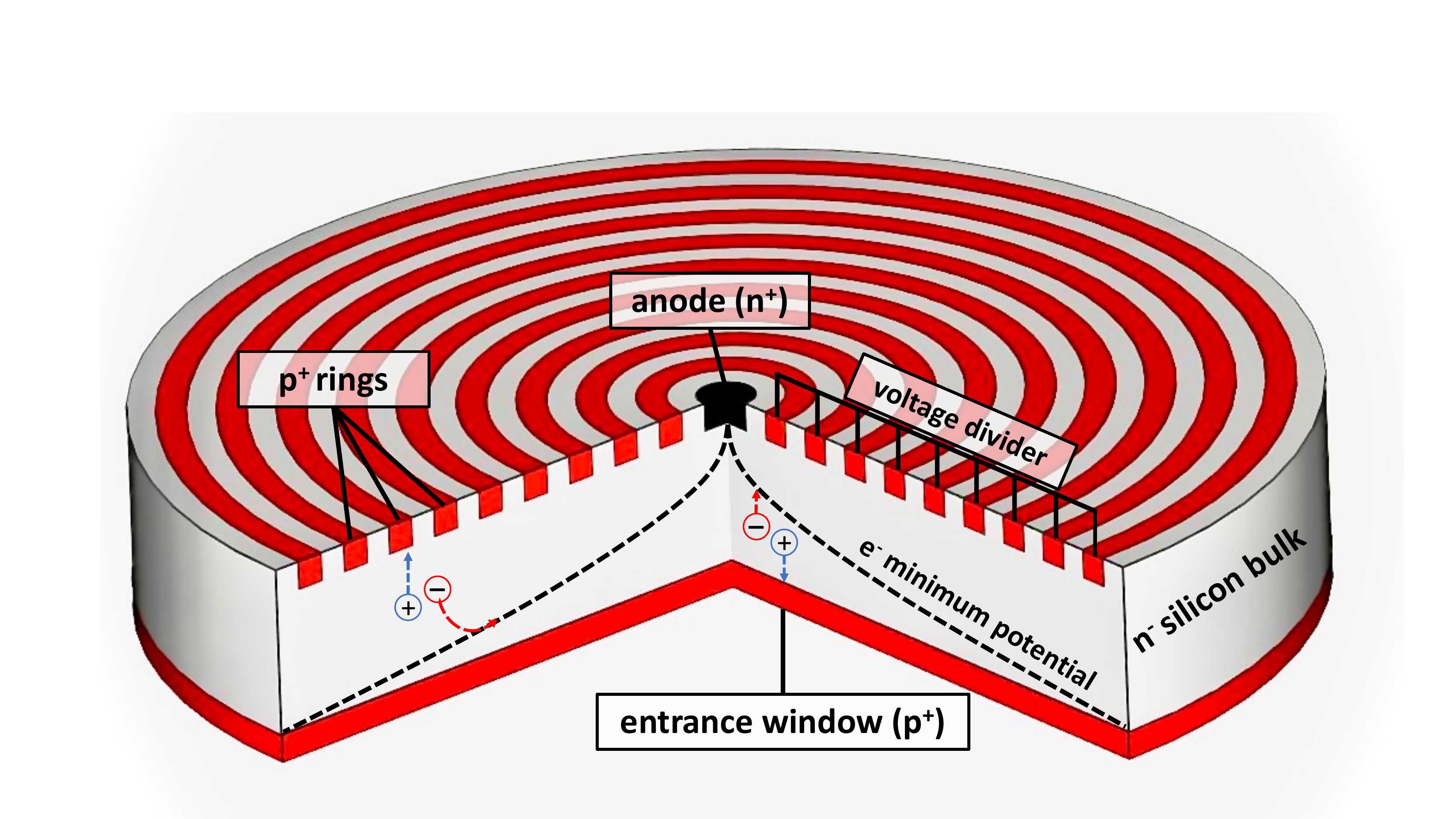}}
\caption{Schematic layout of a cylindrical Silicon Drift Detector. The dotted line represents the diagonal shape of the drift field inside the bulk.}
\label{SDD_layout}
\end{figure}

The working principle of a SDD is based on the to \emph{p-n} diode technology. A large depleted region is created in the silicon bulk, and the electron-hole pairs, generated by the incident radiation, are separated through a reverse polarization field. After the complete depletion of the silicon wafer, a second electric field is superimposed to transport the charges to the collection anode. The SDDs used by SIDDHARTA-2 have a cylindrical shape and are formed by a $n^-$ silicon bulk, with ring-shaped $p^+$ strips on one side and a $p^+$ non-structured layer (entrance window) on the other side to give a homogeneous sensitivity over the entire detector area. The $n^+$ collecting anode is placed in the center of the ring shaped strips (see Figure \ref{SDD_layout}). 
A negative voltage, with respect to the $n^+$ anode, is applied to the $p^+$ entrance window and the $p^+$ ring strips on the opposite side, to fully deplete the \emph{n-type} bulk.
The negative voltage of the $p^+$ rings increases from the ring next to the $n^+$ anode to the outermost one. The voltage of the outermost ring is twice that of the back contact \cite{Fiorini:2000zi}. This voltage configuration generates a ``gutter-like" field \cite{LECHNER1996346} that transports the charges to the collection anode. Figure \ref{SDD_layout} shows the cross section of the SDD and the diagonal shape of the drift field inside the bulk. The electrons generated inside the depleted volume of the detector by the incident radiation will therefore drift to the $n^+$ collecting anode. Instead, the holes are collected by the reverse biased $p^+$ regions.
A junction field-effect transistor (J-FET) designed to work on a fully depleted silicon substrate, can be installed close to the $n^+$ anode, acting as first amplifying stage \cite{LECHNER1996346}. The integration of the transistor onto the SDD ensures the matching of the input capacitance of the J-FET to the output capacitance of the detector. Moreover, a reset mechanism is included to remove the charge accumulated, due to the leakage current and the signals, by opening a path for the electrons from the anode to the clear contact. The reset pulse is performed by applying a positive voltage pulse to the $n^+$ clear contact. \\
The small capacitance of the anode is the main feature of the SDDs. It provides a lower rise time, high amplitude of the output signal, and consequently less electronic noise and high energy and time resolutions. Moreover, the capacitance is independent from the detector's active area, allowing to produce detectors with a large area.

\subsection{The Silicon Drift Detectors for the SIDDHARTA-2 experiment}
\begin{figure}[htbp]
\centering
\includegraphics[width=8 cm]{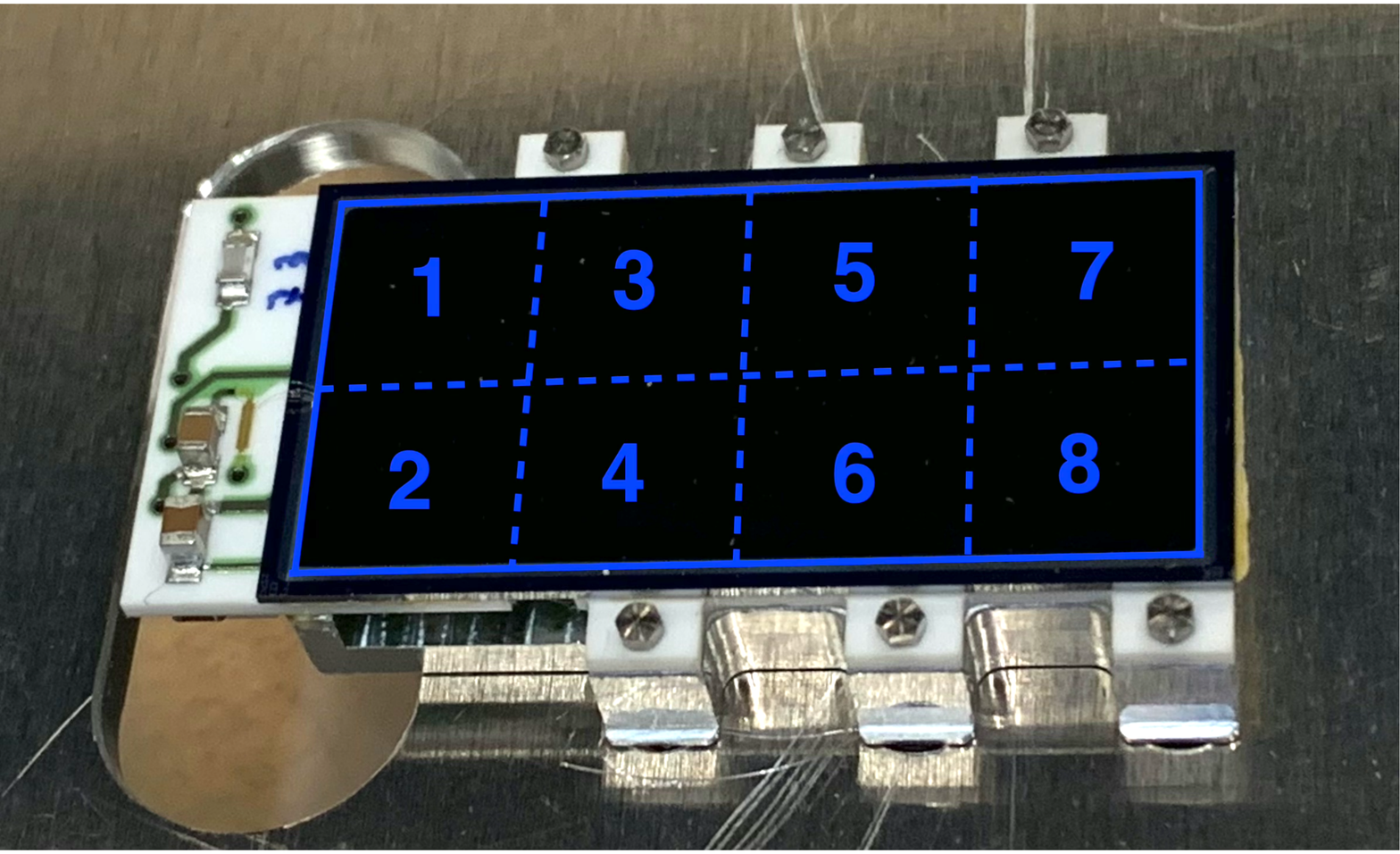}
\caption{An SDD array for the SIDDHARTA-2 experiment.}
\label{SDD_img}
\end{figure}

\begin{figure}[htbp]
\centering
\mbox{\includegraphics[width=15 cm]{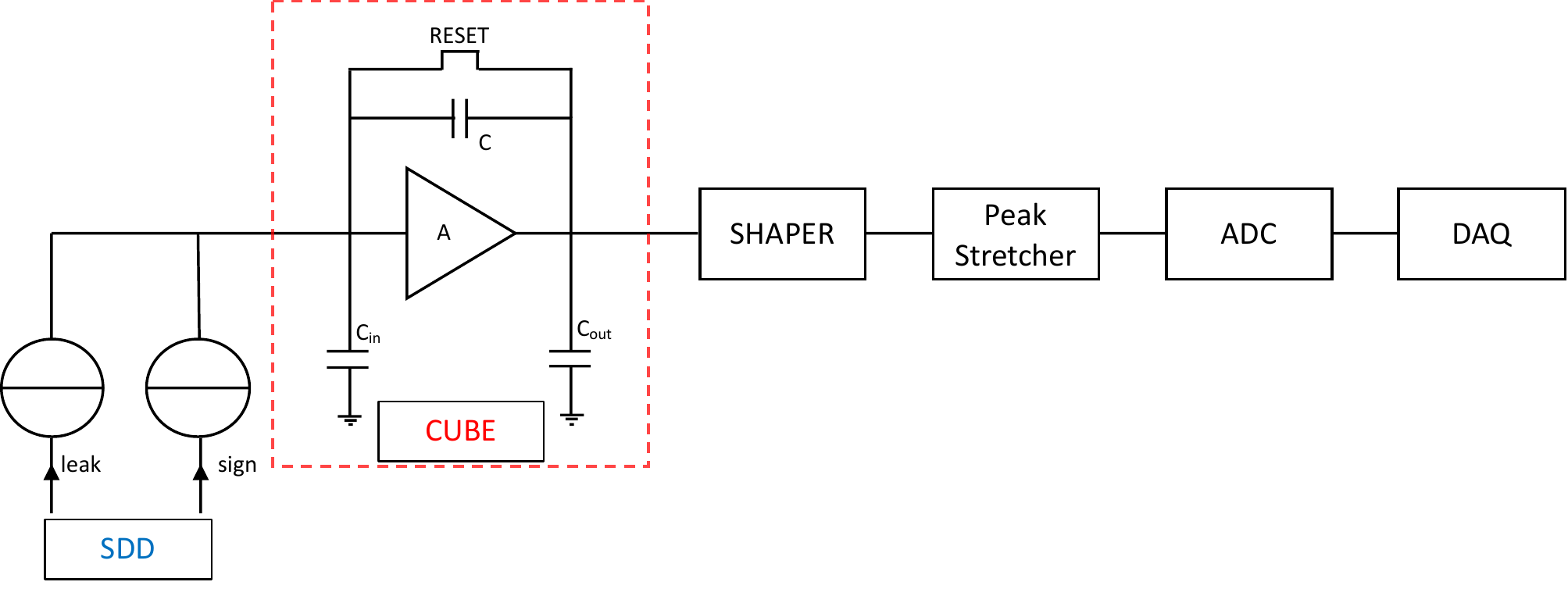}}
\caption{SDD analog chain block diagram.}
\label{anal_chain}
\end{figure}

New monolithic SDD arrays  (see Figure \ref{SDD_img}) have been developed by Fondazione Bruno Kessler (FBK, Italy) in collaboration with Politecnico di Milano (PoliMi, Italy), Istituto Nazionale di Fisica Nucleare - Laboratori Nazionali di Frascati (INFN-LNF, Italy) and Stefan Meyer Institute (SMI, Austria), for the kaonic deuterium measurement at INFN-LNF (SIDDHARTA-2 experiment) and J-PARC (E57 experiment \cite{Curceanu:2019uph}).\\
A monolithic SDDs array consists of eight square SDD cells, each with an active area of 8$\times$8 mm$^2$. The 450 $\mathrm{\mu}$m thick silicon bulk allows to achieve an efficiency of almost 100\% for X-rays in the energy between 5-12 keV, which corresponds to the region of interest for the kaonic deuterium measurement \cite{Curceanu:2019uph}. The individual SDD cells are arranged in a 2$\times$4 array with a 1 mm dead region along the device's borders. The silicon wafer is glued on an alumina ceramic carrier, which provides the polarization voltages, and screwed on an aluminium holder. This high thermal conductive block protects the delicate detector bonding connections and is used to cool down the SDDs to temperatures between 100 K and 150 K. The special ``gear-wheel" structure of the ceramic carrier allows close packing of several SDD arrays, which is essential for optimizing the geometrical X-ray detection efficiency. These SDD arrays belong to an improved technology with an average leakage current of 25 pA$/$cm$^2$ at room temperature \cite{Bertuccio_2015}.
A very significant improvement is the change of the preamplifier system from the J-FET to a complementary metal-oxide semiconductor integrated charge sensing amplifier, named CUBE \cite{6154396}. For each SDD cell, the CUBE is placed on the ceramic carrier as close as possible to the collecting anode and connected to it with a bonding wire. Thanks to the new preamplifier CUBE, the SDDs' performances are stable even when exposed to high and variable charged particle rates \cite{6154396} and a faster drift time is achievable thanks to the possibility to cool down to 100 K, which was not possible with the J-FET, which reach an optimal performance around 170 K (see Table \ref{tab1}).\\
\noindent
The signals are processed by a dedicated front-end electronics and DAQ system. The output of the CUBE is connected to a common ASIC called SFERA (SDDs Front-End Readout ASIC), an Integrated Circuit performing analog shaping and peak detection of the signals. Each SFERA was designed by Politecnico di Milano to read 16 SDDs \cite{Schembari:2016IEE,Quaglia:2016uox}. 
The SFERA main shaper is characterized by a 9th order semi-Gaussian complex-conjugate poles filter with selectable peaking times (from 500 ns to 6 $\mathrm{\mu}$s) and gains, while the fast shaper has a fixed 200 ns peaking time, and is used for pile-up rejection. SFERA allows to acquire both timing and energy information for multiple hits on the SDDs. Figure \ref{anal_chain} shows a schematic representation of the analog chain block diagram corresponding to a single SDD. The signal generated by the incident radiation consists of an electron packet that can be represented as current pulse. It is collected at the anode of the SDD and is amplified by the CUBE which represents the first element of the analog chain. During the data taking, the leakage current of the detector and the electron packets charge the CUBE's capacitor. The output consists of a slow constant slope ramp due to the leakage current, while the collection of an electron packet generates a rapid increase of the charge accumulated on the capacitor. Thus, the collection of an electron packet produces a step on the ramp, this pulse is isolated through a shaper, whose shaping time of 2 $\mathrm{\mu}$s is chosen to minimize the noise. Then, the maximum value of the Gaussian output is processed by the Peak Stretcher circuit and, the voltage signal is converted into a digital signal through an analog-to-digital converter (ADC). Lastly, the full event data information is saved in a file.

\subsection{Silicon Drift Detector's energy response}
The SIDDHARTA-2 experimental apparatus \cite{condmat6040047}, currently installed at the DA$\Phi$NE collider \cite{Zobov:2010zza,Milardi:2018sih} of INFN-LNF, uses 48 SDD arrays, for a total amount of 384 SDDs. Each SDD was characterized before installation in the setup. The laboratory tests show that the energy response of the SDDs is linear within 2-3 eV in the energy range between 4.5-12 keV and the energy resolution is 140 eV at 6.4 keV \cite{Miliucci:2019mdpi}. Given the excellent spectroscopic response of the SIDDHARTA-2 SDD system \cite{Miliucci:2021wbj}, proven during the DA$\Phi$NE beam commissioning phase in the high energy particle background of the collider, the definition of a robust calibration method and a stability check of the SDD system are mandatory to achieve the high precision kaonic deuterium measurement aimed by the SIDDHARTA-2 experiment.

\section{SDD Calibration and Stability}
\subsection{Calibration method}\label{sec_calib}
\begin{figure}[htbp]
\centering
\mbox{\includegraphics[width=15 cm]{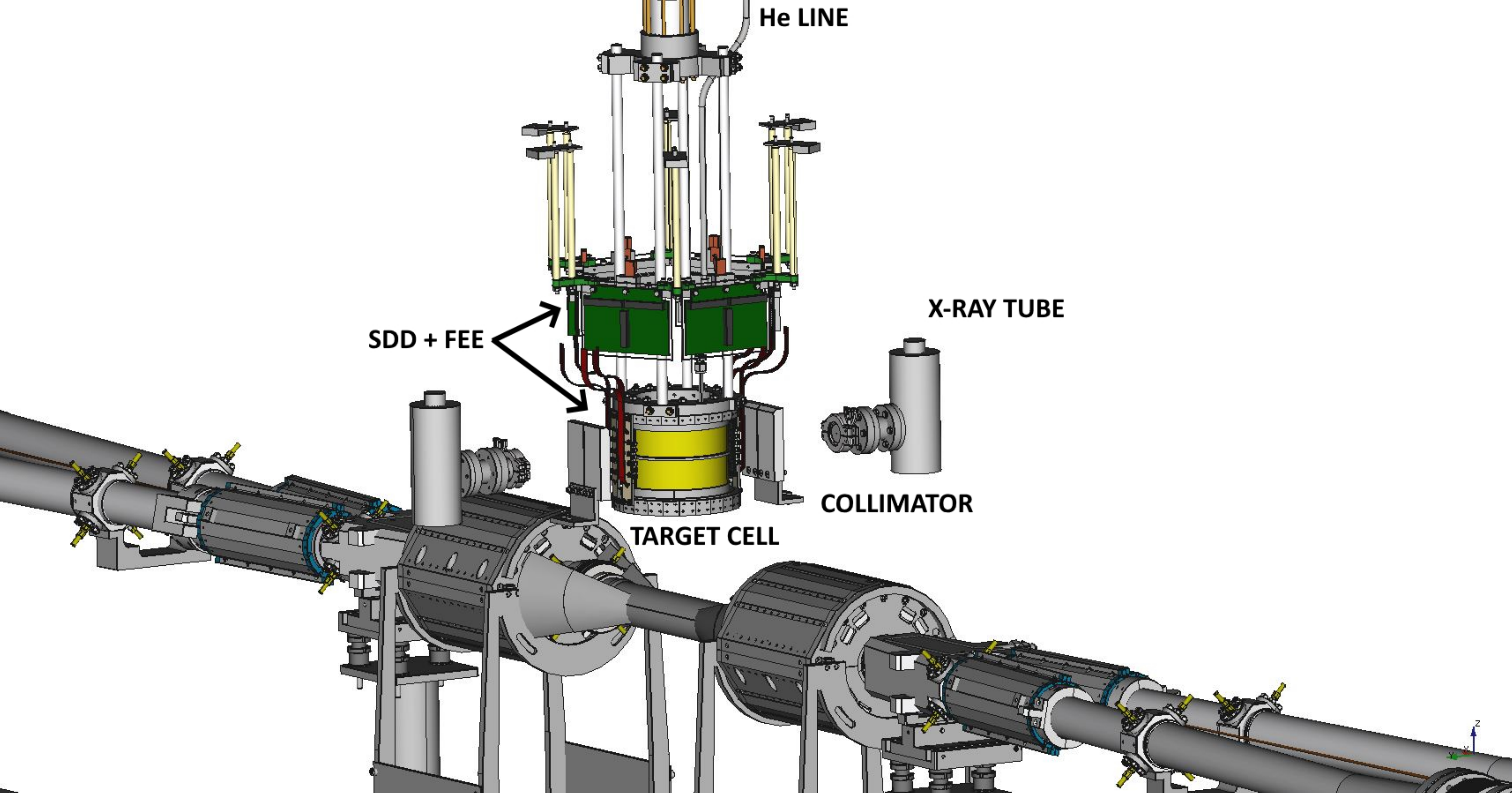}}
\caption{Schematic layout of the SIDDHARTA-2 calibration system. The target cell, the SDDs with the Front-End-Electronic (FEE) and the X-ray tubes are visible above the DA$\Phi$NE beam line.}
\label{setup}
\end{figure}

\begin{figure}[htbp]
\centering
\mbox{\includegraphics[width=7 cm]{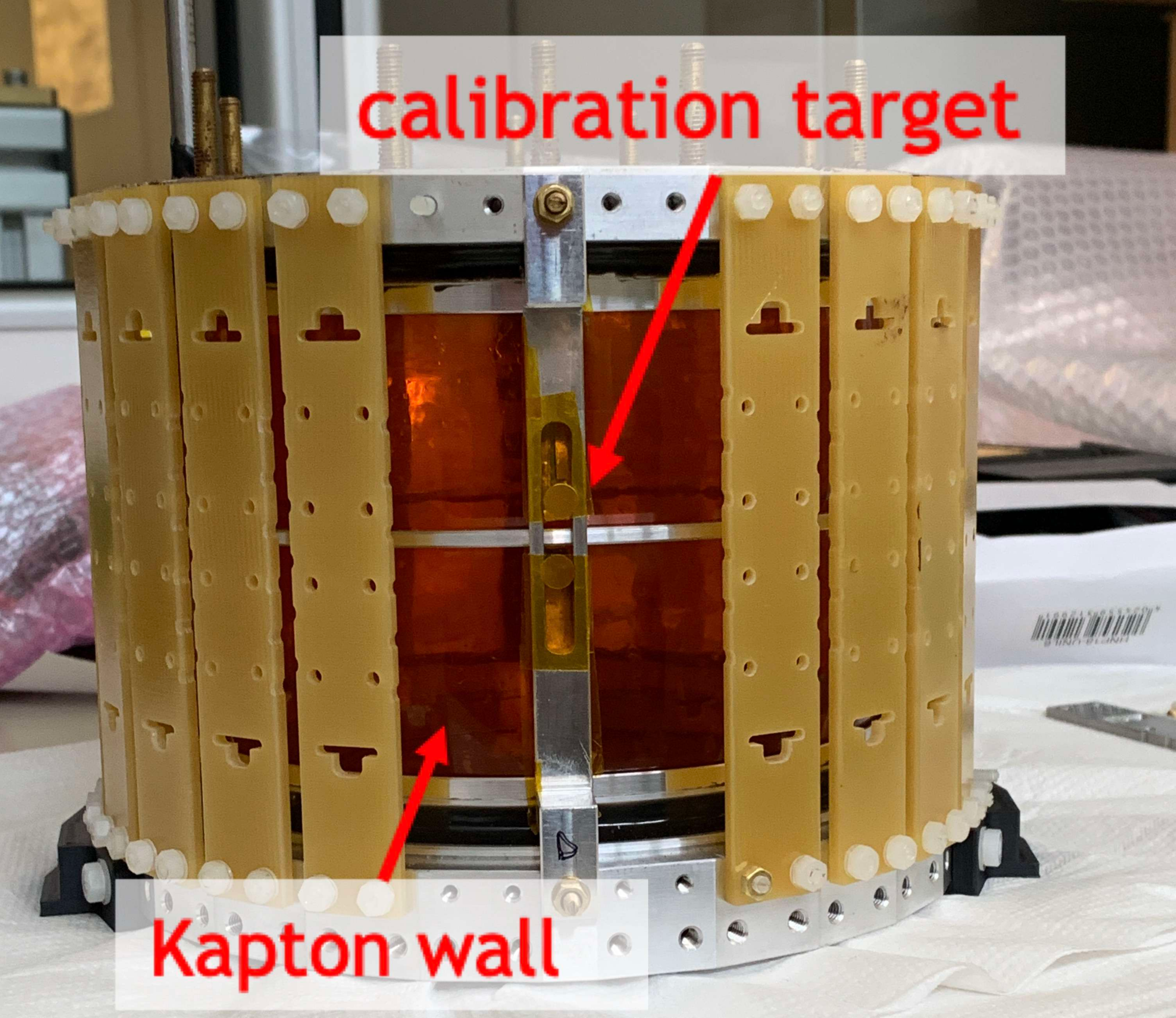}}
\caption{Target cell and multi elements target used for the SDD calibration.}
\label{target}
\end{figure}

\begin{figure}[htbp]
\centering
\includegraphics[width=12 cm]{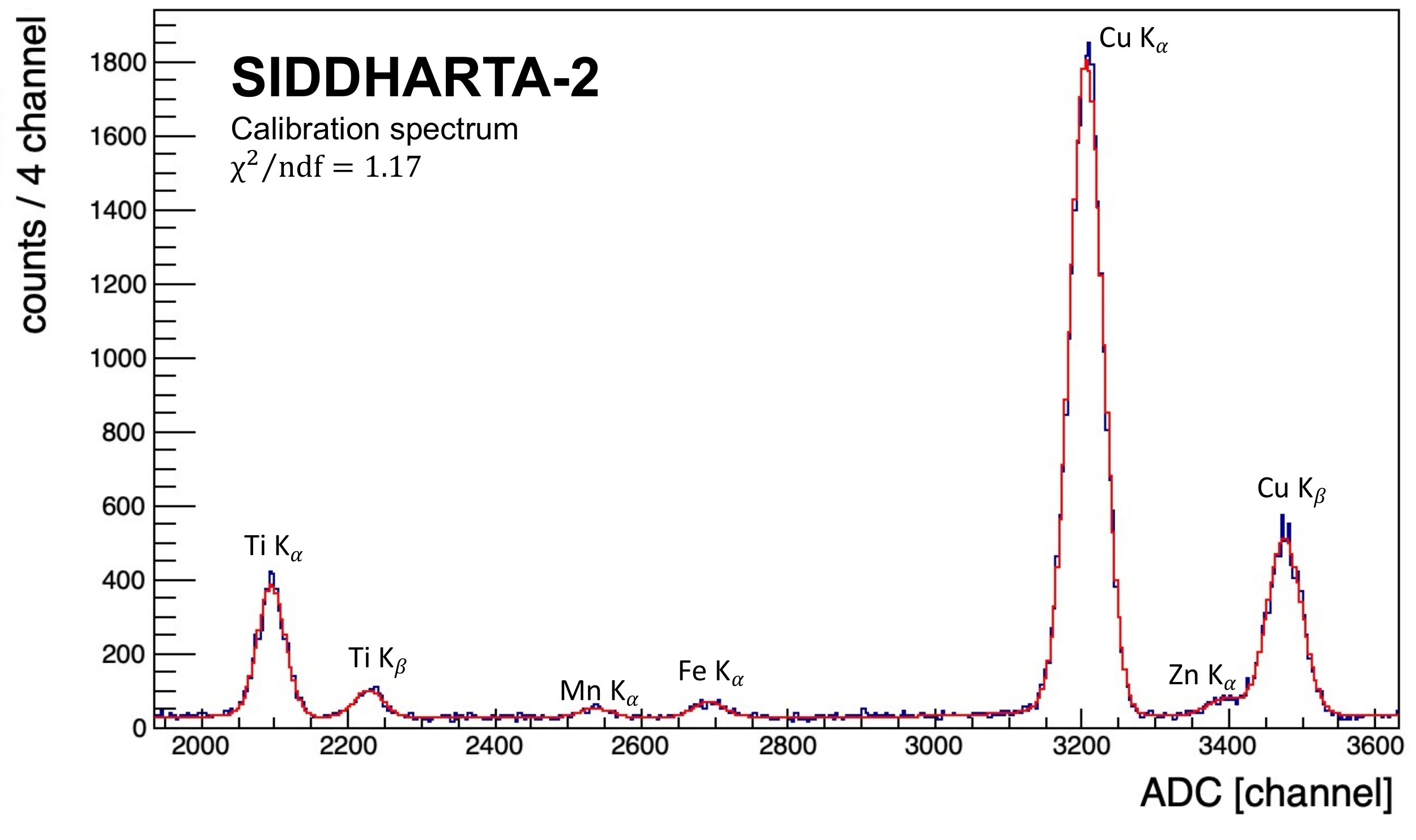}
\includegraphics[width=12 cm]{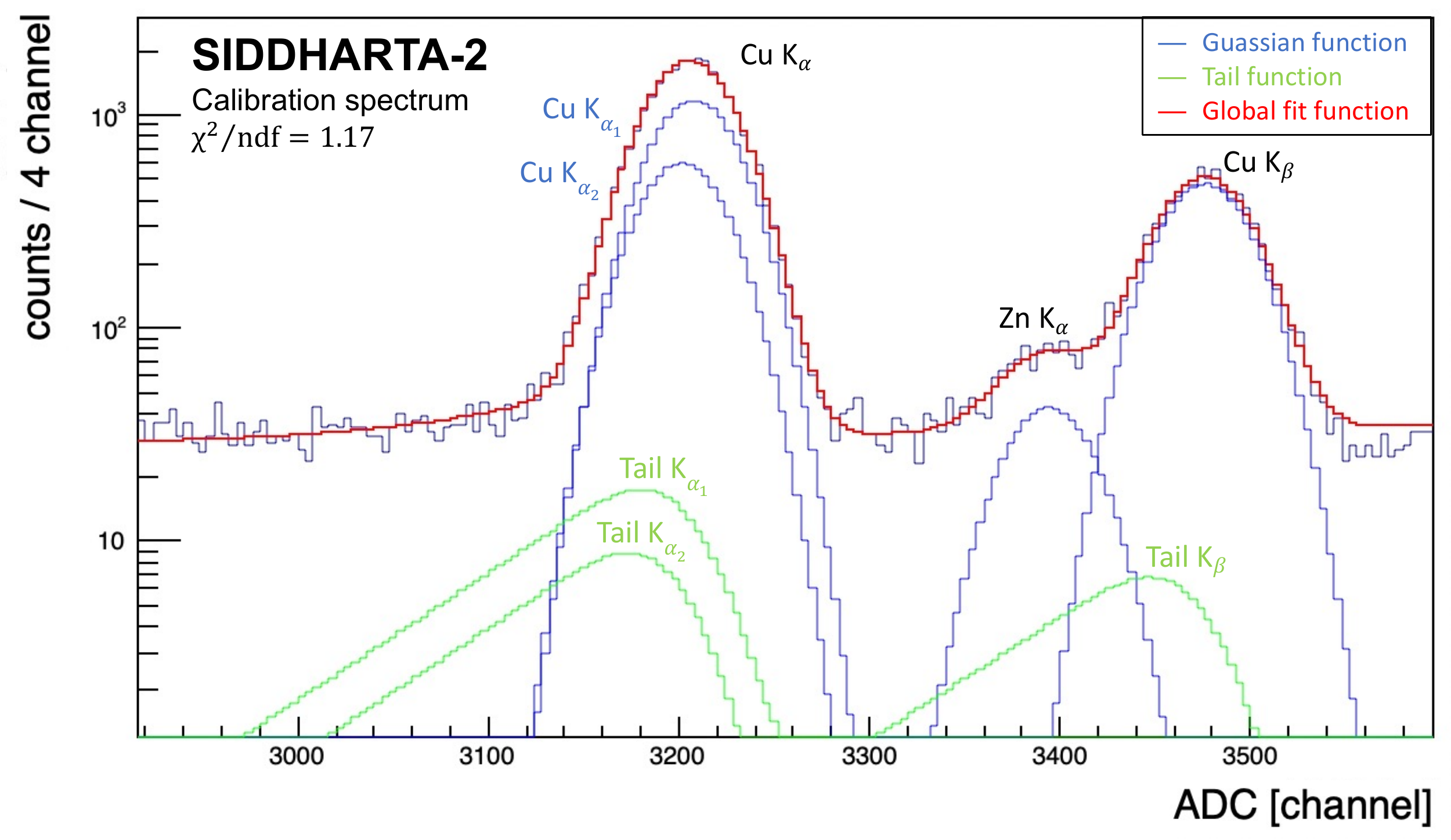}
\caption{Top: typical calibration spectrum for a single SDD in arbitrary units given by ADC. The red line represents the fit function obtained by convolution of a gaussian with an exponential low energy tail for each peak together with a constant function plus an exponential function to reproduce the background shape.
Bottom: highlights of the various contributions to the fit function.}
\label{calib}
\end{figure}

Figure \ref{setup} shows a schematic view of the SIDDHARTA-2 calibration system. The SDDs are placed around the cryogenic target cell made of high purity aluminium structure and 75 $\mathrm{\mu}$m thick Kapton walls.
The energy calibration of the SDDs is performed using two X-ray tubes installed on two sides of the vacuum chamber, and a multi element target made of high purity titanium and copper strips placed on the target cell walls (see Figure \ref{target}). The X-ray tubes induce the fluorescence emission of the target elements and the characteristic K$_\alpha$ and K$_\beta$ transitions are detected by the SDDs. Figure \ref{calib} shows a typical calibration spectrum for a single SDD. The titanium and copper lines are clearly visible together with the Mn K$_\alpha$, Fe K$_\alpha$ and Zn K$_\alpha$ lines produced by the accidental excitation of other components of the setup.\\
The energy response function of the detector is predominantly a Gaussian curve for every fluorescence X-ray peak; however the response has a low energy component due to the incomplete charge collection and electron-hole recombination. Thus, the total peak fit function is formed by two contributions:
\begin{itemize}
\item \emph{Gauss function.} The main contribution to the peak shape. The width of the peak ($\sigma$) is described as function of the Fano Factor ($FF$), the electron-hole pair energy creation ($\epsilon$) and the electronic and thermal noises ($noise$):
         \begin{equation}
         G(x)=\frac{A_G}{\sqrt{2\pi}\sigma} \cdot e^{\frac{-(x-x_0)^2}{2\sigma^2}} \\
         \sigma=\sqrt{FF\cdot\epsilon\cdot E+\frac{noise^2}{2.35^2}} 
         \label{gauss_func}
         \end{equation}

\item \emph{Tail function.} An exponential function to reproduce the incomplete charge collection: 

          \begin{equation}
          T(x)=\frac{A_T}{2\beta\sigma} \cdot e^{\frac{x-x_0}{\beta\sigma}+\frac{1}{2\beta^2}} \cdot\emph{erfc}\left(\frac{x-x_0}{\sqrt{2}\sigma}+\frac{1}{\sqrt{2}\beta}\right)
          \label{tail_func}
          \end{equation}
\end{itemize}  
The constants $A_G$ and $A_T$ are the amplitude of the Gauss and Tail functions, respectively. The $\beta$ parameter is the slope of the tail, while \emph{erfc} is the complementary error function. A constant function plus an exponential are used to reproduce the background shape.\\
\noindent
Only the Ti K$_\alpha$ and Cu K$_\alpha$ peaks are exploited to calibrate the detectors, since they have the highest signal-to-background ratio. The K$_\alpha$ peak is the convolution of two transition lines, K$_{\alpha1}$ and K$_{\alpha2}$. Thus, the calibration peaks were fitted using gaussian and tail functions for each component, to improve the calibration accuracy. Since the detector response is linear \cite{Miliucci:2021wbj}, the energy distance between the K$_{\alpha1}$ and  K$_{\alpha2}$ was set to the reference value as well as was the relative amplitude \cite{xraybooklet}. \\
\noindent
The Fe K$_\alpha$ peak, not included in the determination of calibration parameters, can be used to evaluate the goodness of the calibration itself. After the sum of the calibrated spectra for all the SDDs, a fit was performed to obtain the energy position of the Fe K$_\alpha$ peak. The deviation of the Fe K$_\alpha$ with respect to the tabulated reference value \cite{xraybooklet} is 2 $\pm$ 0.1 eV, which characterizes the accuracy of the calibration method, compatible to the linearity of the system.

\subsection{Stability}
\begin{figure}[htbp]
\centering
\mbox{\includegraphics[width=16 cm]{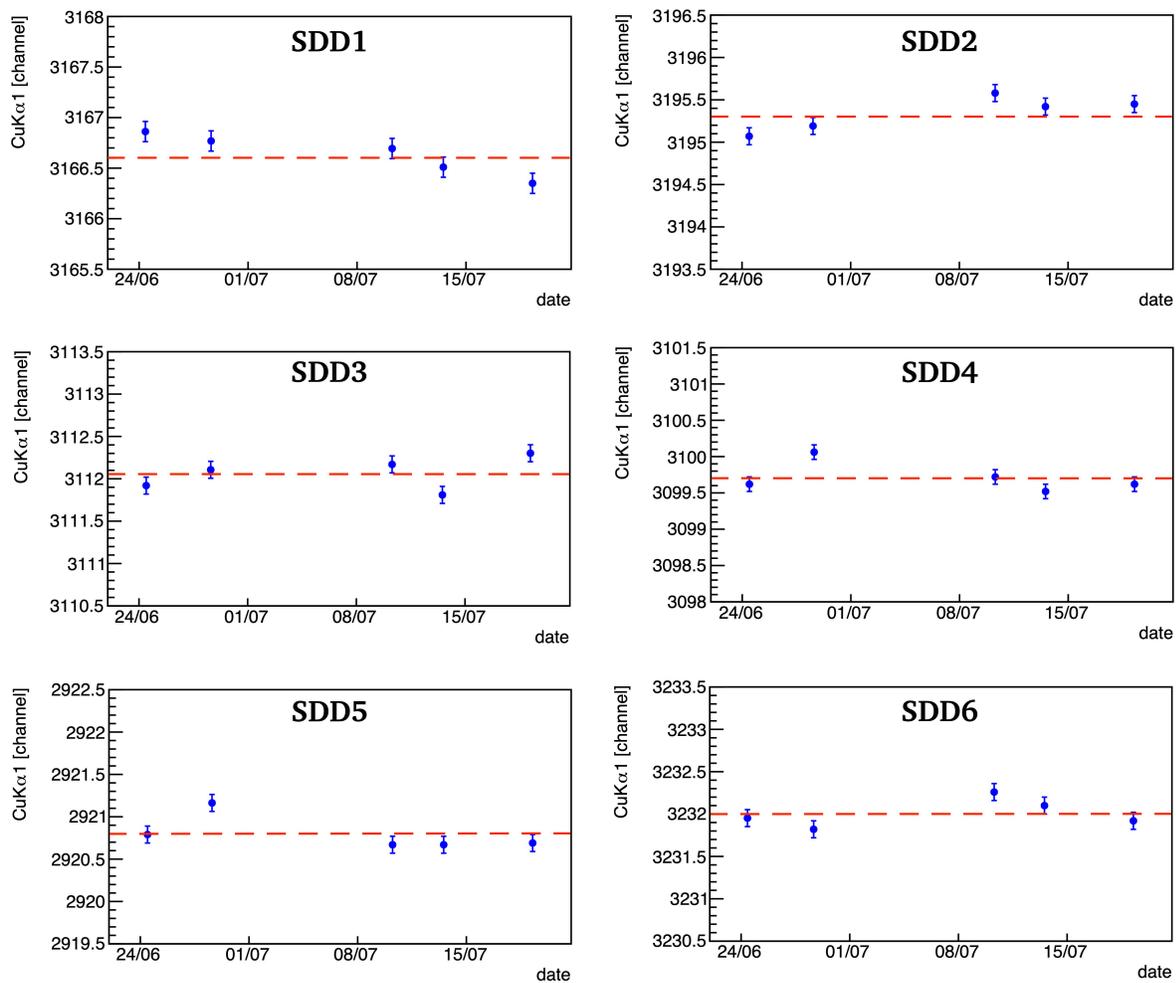}}
\caption{Cu K$_\alpha$ peak position, given in arbitrary units of ADC, as function of time for several SDDs calibration runs.}
\label{stability_time}
\end{figure}

The stability of the SDDs is crucial to perform high precision measurements because it reflects a possible source of systematic error for the SIDDHARTA-2 experiment. Moreover, a stable system needs fewer calibration runs during the kaonic atoms data taking and consequently will increase the experiment's runtime. For these reasons, stability is a fundamental parameter that needs to be monitored carefully.\\
Since the linearity of the SDDs is within 2-3 eV, the stability of the system should be at least at the same level. During the first phase of the SIDDHARTA-2 experiment, which took place from June 2021 to July 2021, we monitored the stability of the SDDs for about one month and an integrated luminosity of 30 pb$^{-1}$. During this period, 5 calibration runs were performed every $\sim$ 6 pb$^{-1}$ of data taken. For each SDD, the calibration method described in paragraph \ref{sec_calib} was used to control the fluctuation over time of the Cu K$_\alpha$ peak position. In Figure \ref{stability_time} the position of Cu K$_\alpha$, given in arbitrary units of ADC, for several SDDs is shown. The fluctuations of the Cu K$_\alpha$ position are about 0.5 channels, which corresponds to $\sim$ 1.5 eV. This result proves that the stability of the SIDDHARTA-2 SDD system is suitable to perform high precision kaonic atoms measurements, when calibration runs every $\sim$ 6 pb$^{-1}$ of integrated luminosity are performed.\\

\begin{figure}[htbp]
\centering
\mbox{\includegraphics[width=13 cm]{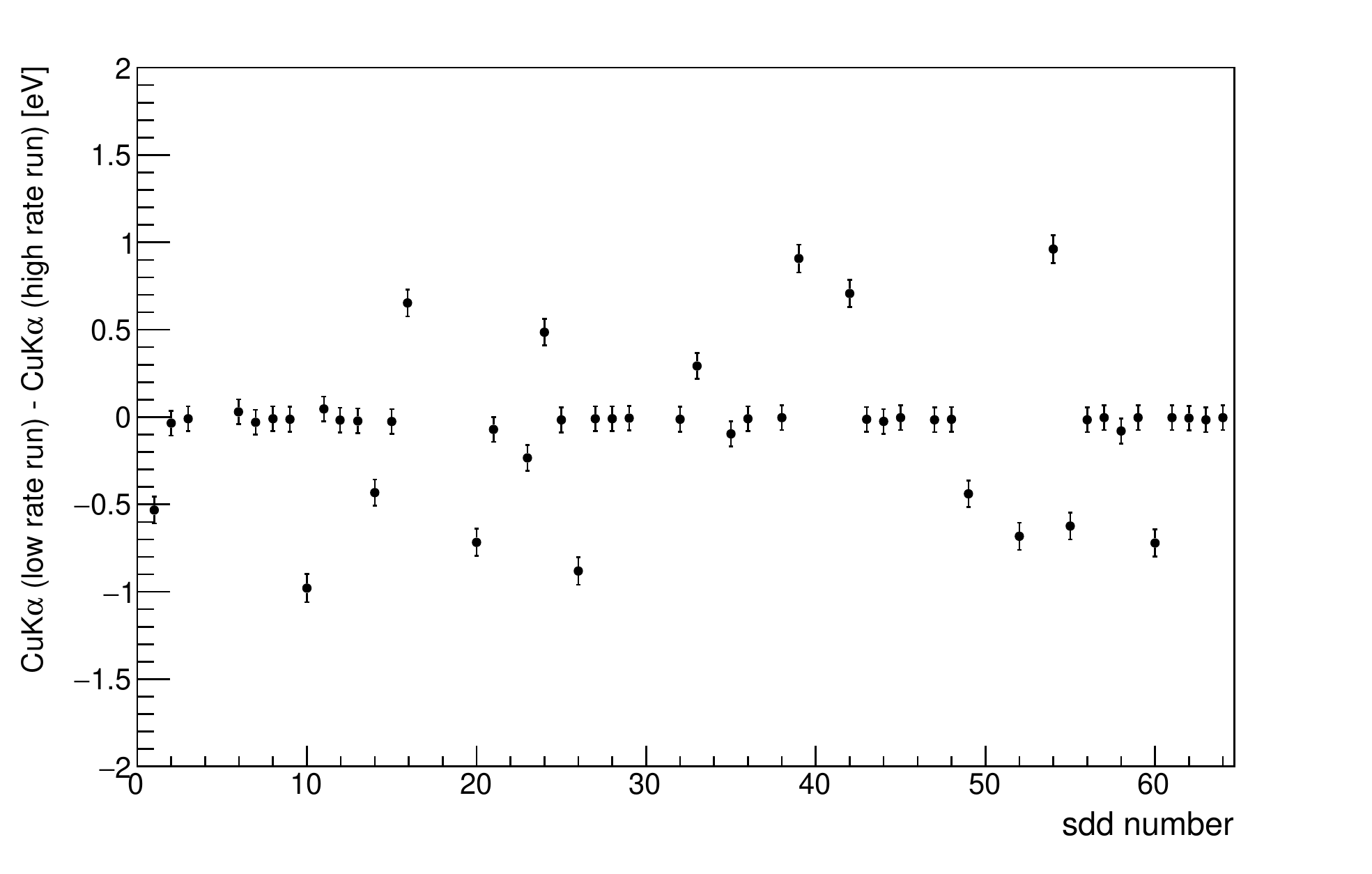}}
\caption{Difference in Cu K$_\alpha$ peak position for high rate and low rate runs for several SDDs.}
\label{stability_rate}
\end{figure}

\noindent
It is also important that the energy response of the detectors is stable and independent of the counting rate, since during the beam injection the SDD rate can increase up to several hundreds of Hz, and then gradually decrease to a few tens of Hz. For this, we performed two types of calibration runs varying the voltage and current parameters of the X-ray tubes in order to obtain two different counting rates (60 Hz and 600 Hz). Figure \ref{stability_rate} shows the difference in Cu K$_\alpha$ peak position for high and low counting rates for several SDDs. For most detectors the difference is almost zero, for the others the fluctuation is compatible with the accuracy of the calibration. In conclusion, the energy response of the SDD system can be considered independent of the counting rate, within 1 eV.

\section{Conclusions}
Precision X-ray spectroscopy of light kaonic atoms requires state-of-the-art silicon detectors to meet the demands of high energy and time resolutions, as well as radiation hardness. In this context, innovative SDDs have been developed by the SIDDHARTA-2 collaboration to perform the challenging kaonic deuterium measurement. The SDD calibration method was tested during the first phase of the SIDDHARTA-2 run and its performance is described in this paper. The design of the calibration system and the detailed analysis of the spectra were driven by the requirement of an accuracy of the calibration at the level of 2-3 eV to be consistent with the SDDs' linearity. The optimized method has been applied also to define the long-term stability and the independence from the rate of the SDDs spectroscopy response. The results presented in this paper show that the SIDDHARTA-2 SDD system performances are suitable to accomplish the challenging kaonic deuterium measurement with a systematic error at the level of a few eV, keeping stable its high precision X-ray spectroscopy response over the whole data taking  period.

\ack
The authors acknowledge C. Capoccia from INFN-LNF and H. Schneider, L. Stohwasser, and D. Pristauz Telsnigg from Stefan-Meyer-Institut f\"ur Subatomare Physik for their fundamental contribution in designing and building the SIDDHARTA-2 setup. We thank as well the DA$\Phi$NE staff for the excellent working conditions and permanent support.
Part of this work was supported by the Austrian Science Fund (FWF): P24756-N20 and P33037-N; the Croatian Science Foundation under the project IP-2018-01-8570; EU STRONG-2020 project (grant agreement No.824093), the EU Horizon 2020 project under the MSCA G.A. 754496, the Polish Ministry of Science and Higher Education grant No. 7150/E-338/M/2018 and the Foundational Questions Institute and Fetzer Franklin Fund, a donor advised fund of Silicon Valley Community Foundation (Grant No. FQXi-RFP-CPW-200)

\section*{References}
\bibliography{iopart-num.bib}

\end{document}